\documentclass[twocolumn]{aastex701} 
\usepackage{textcomp}
\usepackage{array}
\usepackage{tabularx}

\usepackage{siunitx}

\usepackage{multirow}
\usepackage{color} 
\usepackage{makecell}
\usepackage{listings}
\usepackage{xcolor}
\usepackage{graphicx}
\graphicspath{{figures/}}

\begin{document}

\title{Hot Molecular Hydrogen in Galaxies at Redshift $\sim$ 0.2--1.5 Revealed by JWST}
\author[0000-0002-4765-1500]{Jie Chen}
\affiliation{Department of Astronomy, School of Physics, Peking University, Beijing 100871, People's Republic of China}
\affiliation{Kavli Institute for Astronomy and Astrophysics, Peking University, Beijing 100871, People's Republic of China}
\email{jiechen@stu.pku.edu.cn}

\author[0000-0003-4176-6486]{Linhua Jiang}
\affiliation{Department of Astronomy, School of Physics, Peking University, Beijing 100871, People's Republic of China}
\affiliation{Kavli Institute for Astronomy and Astrophysics, Peking University, Beijing 100871, People's Republic of China}
\email{jiangKIAA@pku.edu.cn}

\author[0000-0003-0964-7188]{Shuqi Fu}
\affiliation{Department of Astronomy, School of Physics, Peking University, Beijing 100871, People's Republic of China}
\affiliation{Kavli Institute for Astronomy and Astrophysics, Peking University, Beijing 100871, People's Republic of China}
\affiliation{Jodrell Bank Centre for Astrophysics, University of Manchester, Oxford Road, Manchester, UK}
\email{fushuqi@stu.pku.edu.cn}

\author[0000-0002-2420-5022]{Zijian Zhang}
\affiliation{Department of Astronomy, School of Physics, Peking University, Beijing 100871, People's Republic of China}
\affiliation{Kavli Institute for Astronomy and Astrophysics, Peking University, Beijing 100871, People's Republic of China}
\email{zjz.kiaa@stu.pku.edu.cn}

\author[0009-0003-6747-2221]{Danyang Jiang}
\affiliation{Department of Astronomy, School of Physics, Peking University, Beijing 100871, People's Republic of China}
\affiliation{Kavli Institute for Astronomy and Astrophysics, Peking University, Beijing 100871, People's Republic of China}
\email{jiangdy@stu.pku.edu.cn}

\author{Jing Wang}
\affiliation{Department of Astronomy, School of Physics, Peking University, Beijing 100871, People's Republic of China}
\affiliation{Kavli Institute for Astronomy and Astrophysics, Peking University, Beijing 100871, People's Republic of China}
\email{}

\author[0000-0003-4956-5742]{Ran Wang}
\affiliation{Department of Astronomy, School of Physics, Peking University, Beijing 100871, People's Republic of China}
\affiliation{Kavli Institute for Astronomy and Astrophysics, Peking University, Beijing 100871, People's Republic of China}
\email{}

\author{Darshan Kakkad}
\affiliation{Centre for Astrophysics Research, Department of Physics, Astronomy and Mathematics, University of Hertfordshire, Hatfield, AL10 9AB, UK}
\email{}

\begin{abstract}

Molecular hydrogen (H$_2$) is the fundamental component of the molecular gas reservoir that fuels star formation and black hole growth. While prominent hot H$_2$ emission ($\gtrsim 1000$\,K) has been explored in nearby galaxies, its excitation mechanisms at higher redshifts remain poorly constrained. We present a systematic study of hot molecular gas traced via ro-vibrational H$_2$ emitters at $z \sim$ 0.2--1.5 using JWST slitless spectroscopy. Exploiting JWST's great sensitivity to near-infrared H$_2$ ro-vibrational transitions such as H$_2$ 1-0 S(1), we build a sample of 144 H$_2$-luminous galaxies and characterize their multi-wavelength properties. About 96.5\% and 80\% of these galaxies are detected in the radio and far-infrared (FIR) bands, respectively, and their hot H$_2$ emission strongly correlates with the radio and FIR radiation. Based on radio excess, X-ray emission, and optical diagnostics, 62 sources (43\%) exhibit significant AGN activity. The H$_2$ line luminosities of the sample show a sublinear dependence on radio luminosity, while the relation between the H$_2$ luminosities and total infrared luminosities is approximately linear after excluding the radio-excess sources. These scaling relations suggest that star formation provides a baseline contribution to hot H$_2$ excitation, while radio-excess sources require an additional contribution from AGN-related processes such as jet- or wind-driven shocks. These results reveal a multi-channel excitation scenario and shed new light on the interplay between star formation, AGN activity, and the interstellar medium during galaxy evolution.

\end{abstract}

\keywords{Emission line galaxies (459);  Molecular gas (1073); Surveys (1671); James Webb Space Telescope (2291);}


\section{Introduction} \label{sec:intro}

Molecular hydrogen (H$_2$) is the most abundant molecule and constitutes the dominant reservoir of cold gas that fuels star formation and black hole accretion in galaxies. Cold H$_2$ is difficult to detect directly because H$_2$ lacks a permanent dipole moment and its rotational transitions are weak quadrupole transitions. The upper levels of H$_2$ ro-vibrational transitions correspond to relatively high excitation temperatures ($T_{\rm ex}\gtrsim500\,$K), which are much higher than those of typical cold molecular clouds. Therefore, cold H$_2$ is usually traced indirectly through CO and dust emission \citep{2013ARA&A..51..105C,2022ARA&A..60..319S}, while warm and hot H$_2$ ($\sim$200–4000\,K) in excited environments can be traced through near-infrared (NIR) ro-vibrational transitions and far-infrared (FIR) pure rotational transitions \citep{1987ApJ...322..412B,1989ApJ...338..178E,2006A&A...457...61R}.
These H$_2$ emission lines provide a powerful diagnostic of the physical conditions, excitation mechanisms, and energy dissipation processes in the interstellar medium (ISM), including ultraviolet (UV) heating in photodissociation regions (PDRs), X-ray irradiation near active galactic nuclei (AGN), and shock heating from stellar feedback or galaxy interactions \citep[e.g.,][]{1994ApJ...427..777M,2007ApJ...669..959R,dasyra11,rupke13}.

The H$_2$ ro-vibrational and pure-rotational emission have been observed in local galaxies and AGNs with ground-based IR telescopes and space FIR telescopes \citep[e.g.,][]{rosario19}. These H$_2$ emissions were found to be particularly prominent in luminous and ultra-luminous infrared galaxies (LIRGs and ULIRGs), which represent extreme environments where intense star formation and AGN activity are deeply embedded in dusty, gas-rich systems \citep{2008ApJ...679..156H}. Palomar long-slit infrared spectrograph observations of 33 ULIRG samples from the Infrared Astronomical Satellite (IRAS) in the local universe have shown that H$_2$ ro-vibrational emission is detected in essentially all nuclear regions, with relatively uniform line ratios and excitation temperatures \citep{2001AJ....121...97M, Burston2001, Davies2003}, consistent with hot H$_2$ thermal excitation assisted by FUV irradiation arising in dense PDRs. \cite{Dannerbauer2005} strengthened the view that these (U)LIRGs are predominantly powered by starbursts and highlighted [Fe II] as a key complement to H$_2$ for tracing feedback and shocks. Only a small fraction of these (U)LIRGs showed AGN signatures, such as broad hydrogen recombination lines or the [Si\,VI] coronal line. 

Integral-field spectroscopy (IFS) over the past decade has revealed that hot H$_2$ emission in (U)LIRGs is typically centrally concentrated, but exhibits complex two-dimensional excitation structures on kpc to sub-kpc scales \citep[e.g.,][]{2012A&A...546A..64P}. Spatially resolved studies demonstrate that different excitation mechanisms coexist within individual galaxies, with contributions from AGN, star-forming regions, and shocks \citep[e.g.,][]{Colina2015,riffel21}. These physical processes can be empirically disentangled using NIR diagnostic line ratios, such as H$_2$/Br$\gamma$ and [Fe II]/Br$\gamma$ for mapping the multiphase interstellar medium in (U)LIRGs \citep{davies09, storchi-bergmann09, friedrich10, ramos-almeida17}.

Both the long-slit K-band spectroscopy and IFS observations of ULIRG nuclei demonstrated that the low-vibrational lines, e.g., H$_2$ 1-0 S(1), are predominantly thermally excited, with typical excitation temperatures of $\sim$1000–2500 K \citep[e.g.,][]{2001AJ....121...97M, Burston2001, Davies2003}. Non-thermal UV fluorescence primarily affects higher vibrational levels, e.g., H$_2$ 2–1 and 3–2 lines. In addition, comparisons between molecular and ionized gas tracers reveal systematic differences in their spatial distributions and kinematics. The hot H$_2$ is generally more confined to the galaxy plane and often follows ordered rotational motion, whereas ionized gas (e.g., traced by recombination lines) tends to exhibit more extended or collimated structures and shows stronger signatures of non-circular motions \citep{2012A&A...546A..64P}. This dichotomy highlights the distinct physical conditions and dynamical roles of molecular and ionized gas phases in (U)LIRGs.

Beyond individual systems, high-resolution studies have begun to reveal statistical trends linking hot H$_2$ properties to galaxy evolution stage and power sources. In particular, AO-assisted integral-field observations \citep[e.g., KOALA in ][]{2019ApJ...871..166U} showed that shocked molecular gas becomes more prevalent in the most infrared-luminous systems and in advanced merger stages, suggesting that dynamical processes such as interactions, gas inflows, and feedback play a key role in enhancing H$_2$ emission. This trend is further supported by detailed case studies of archetypal mergers such as NGC 6240, where velocity-resolved mapping of H$_2$/Br$\gamma$ has been used to disentangle shock- and AGN-heated components within the complex nuclear environment \citep{Lutz2003A&A...409..867L, Ilha2016Ap&SS.361..178I, Carlsen2025arXiv250617584C}. Together, these results point toward a scenario in which AGN activity and merger-driven processes not only regulate the excitation of hot molecular gas but may also systematically boost its luminosity and prominence in the most extreme systems.

The advent of the James Webb Space Telescope (JWST) is now extending NIR H$_2$ studies to much higher redshifts and fainter luminosities, opening a new window on the role of hot molecular gas in galaxy evolution from nearby to the peak epoch of star formation. JWST’s unprecedented sensitivity in the infrared allows direct detection of H$_2$ ro-vibrational lines, including 1–0 S(1), in distant galaxies that were previously inaccessible. Recent observations have reported detections of H$_2$ emission at z $\sim$ 2 in one X-ray AGN by JWST \citep{Kakkad2025}. \cite{2023Kristensen} also made the shock prediction of the H$_2$ spectrum in the JWST era. These observations and theories provide new constraints on excitation mechanisms, gas energetics, and feedback processes at high redshift, and enable direct comparisons with well-studied local (U)LIRGs.

In this work, we constructed a sample of 144 H$_2$-luminous galaxies at redshift $\rm{z}\sim$ 0.2--1.5 using the JWST COSMOS-3D F444W slitless grism survey. The sample is selected based on robust detections of the NIR ro-vibrational transitions, e.g., H$_2$ 1–0 S(1), H$_2$ 1–0 Q(1), and H$_2$ 1–0 Q(3). This large sample enables a systematic study of hot molecular gas across a statistically significant population. We present a systematic study of these ro-vibrational H$_2$ emitters, including the radio/X-ray detection, the construction of the spectral energy distribution (SED) from UV to radio, and any AGN activity in these samples.

This paper is formatted as follows: Section~\ref{sec:data} describes the slitless spectra, photometry data, and optical spectra used in this work. Then we describe the H$_2$ sample selection. Section~\ref{sec:H02} describes the H$_2$ sample properties from multi-wavelength SEDs, radio and IR luminosities, H$_2$ excitation temperatures, and star formation rates (SFRs). Section~\ref{sec:discussion} discusses AGN activities with NIR and optical diagnostics. Section~\ref{sec:Summary} summarizes this study. We adopt a cosmology with $H_0 = 70$\,km\,s$^{-1}$\,Mpc\,$^{-1}$, $\Omega_{\rm{M}} = 0.27$ and $\Omega_\Lambda = 0.73$.

\section{Data reduction and Sample Selection} \label{sec:data}

\subsection{Grism Spectroscopy}

Our grism spectroscopy is primarily based on the JWST COSMOS-3D program. COSMOS-3D is a 268-hour JWST Cycle 3 Large Program (GO-5893) that observes 0.33 $\deg^2$ in the COSMOS field with NIRCam Wide Field Slitless Spectroscopy (WFSS) and broad-band NIRCam and MIRI imaging (Kakiichi et al. in preparation). NIRCam WFSS observations use the row-direction grism (Grism R) combined with the wide-band filter F444W. With its large areal coverage, COSMOS-3D yields a large sample of emission-line sources from the local Universe to high redshift that cover strong lines such as Br$\alpha$, Br$\beta$, Pa$\alpha$, Pa$\beta$, He I, [S III], H$\alpha$, H$\beta$, and [O III] \citep[][Wang et al. in preparation]{Meyer2025, Lin2026,2026arXiv261013920F}. The spectral resolution of WFSS (R$\sim$1600) enables us to identify the weak NIR ro-vibrational H$_2$ transitions.

The WFSS data were reduced following the standard procedure with public code and calibration files\footnote{\url{https://github.com/fengwusun/nircam_grism}} \citep{Sun_2023}. We first processed the NIRCam WFSS data using the standard JWST Stage-1 calibration pipeline (version \verb|1.13.4|) and the calibration reference file \verb|jwst_1364.pmap|. A flat-field correction was applied using flat-field data obtained with the same filter and detector. For each exposure, we constructed and subtracted a super-sky background using all flat-fielded exposures taken with the same filter+pupil+module combination. We then performed an additional 2D sky-background subtraction using  Source Extractor \citep[SExtractor;][]{Bertin_1996} to further remove the residual. We estimated the background from a $2.5\sigma$-clipped image using the mode estimator $(2.5 \times \mathrm{median}) - (1.5 \times \mathrm{mean})$, which reduces sensitivity to source crowding. When 
$(\mathrm{mean} - \mathrm{median})/\mathrm{STD} > 0.3$,
where $\mathrm{STD}$ is the standard deviation, we considered the mode estimator unreliable and instead used a simple median.
Finally, we refined the astrometry of the WFSS data by matching the simultaneously obtained short-wavelength images to the COSMOS2025 source catalog \citep{COSMOS2025_cat}, which was registered to the HST reference frame.

\subsection{Multi-band Photometry}
\begin{figure*}

\centering
\includegraphics[width=1\textwidth]{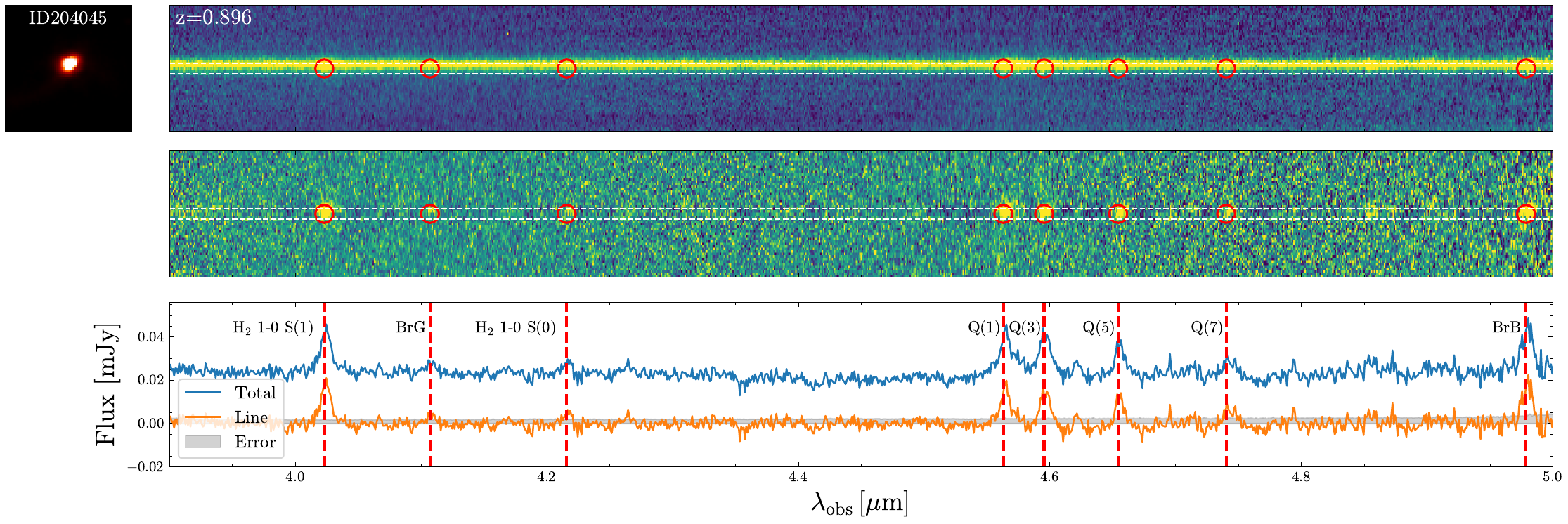}
\caption{Example of an H$_2$ emitter (ID 204045). The upper-left image is a cutout of the direct image in the F444W band. The top and middle panels show the combined 2D spectrum and the continuum-subtracted spectrum, respectively. We mark the detected emission lines with red circles. The bottom panels show the corresponding 1D spectra of the same object.}
\label{fig:ID204045}
\end{figure*}
We used a complete set of JWST/NIRCam images available in the COSMOS field, including the F115W, F150W, F200W, F277W, F356W, and F444W bands. The data were primarily obtained from programs 1635, 1727, 1810, 1837, 1933, 2321, 2514, 3990, 5398, 5893, 6368, 6434, and 6585. The images were reduced using the JWST Calibration Pipeline (v1.17.0), the CEERS NIRCam reduction scripts \citep{Bagley2023}, together with our own custom procedures. We adopted the final JWST photometric catalog from \cite{2026arXiv261013920F}.

For the UV to NIR coverage, we used the CFHT $U$-band images from the MUSUBI (MegaCam Ultra-deep Survey: u*-band Imaging) survey \cite{2022ApJS..260...54W}, Subaru/HSC ultradeep imaging in the $g$, $r$, $i$, $z$, and $y$ bands \citep{Aihara2022}, and Hubble Space Telescope (HST) F814W imaging \citep{Koekemoer2007}, together with the JWST/NIRCam data. For the JWST, HST, and CFHT images, we constructed empirical PSF models for each band using NIRCam grism PSFEx \citep{Bertin2011}. Photometry was performed on PSF-matched images using SExtractor \citep{Bertin_1996} with Kron-aperture corrections. For the HSC data, we derived template-fitting deblended photometry using T-PHOT v2.0 \citep{Merlin2016} to measure total fluxes in each band. Detailed information can be found in \cite{2026arXiv261013920F}.

For the mid-infrared (MIR) data, we used Spitzer/IRAC channels 1–4 and MIPS 24\,$\mu$m imaging from the Spitzer Large Area Survey with Hyper Suprime-Cam \citep[SPLASH;][]{2014ApJSteinhardt} and the COSMOS-Spitzer program \citep[PI: D. Sanders;][]{2009ApJLe}. For the FIR coverage, we used Herschel/PACS 100 and 160\,$\mu$m imaging from PEP \citep[PI: D. Lutz;][]{pep}, together with SPIRE 250, 350, and 500\,$\mu$m nested maps from the Herschel Multi-tiered Extragalactic Survey \citep[HerMES; PI: S. Oliver;][]{hermes}. For the submillimeter and millimeter data, we included SCUBA-2 850,$\mu$m imaging from the S2CLS program \citep{2017Geach}, AzTEC 1.1 mm nested maps \citep{2011Aretxaga}, and MAMBO 1.2 mm imaging \citep{2007ApJSBertoldi}. 

For the radio coverage, we used observations from the VLA-COSMOS 1.4 GHz Large Project, the VLA 3 GHz Survey, the COSMOS-XS survey, the MeerKAT MIGHTEE Survey \citep{vla1.4ghz,vla3ghz,2021ApJvan,2025MNRAS.536.2187H}, as well as MHz-frequency observations from superMIGHTEE \citep{2025ApJ...991....9L}. For the X-ray data, we used imaging from the XMM-Newton Wide-Field Survey and the Chandra COSMOS Legacy Survey \citep{2010Brusa,2016Marchesi}. All these surveys fully cover the COSMOS-3D footprint.

\subsection{H$_2$ Emitter Sample Construction}

We based our parent emission-line catalog from \cite{2026arXiv261013920F} on recent analyses of NIRCam slitless (WFSS) spectroscopy \citep[e.g.,][]{Wang_ASPIRE_2023, Kashino_2023, Fu_2025}. Here we briefly summarize the catalog construction. We adopted a hybrid detection and identification framework optimized for both narrow and broad features. Emission lines were first identified in continuum-subtracted 1D and 2D spectra using matched filtering and signal-to-noise thresholds, with consistency enforced between spectral and spatial detections. To improve sensitivity to broad and low–surface-brightness features that may be suppressed by continuum filtering, an additional 1D-based detection pass is performed with a larger-scale median filter. Detected features were then modeled with single- and double-Gaussian profiles, and classified based on their widths to distinguish narrow lines, broad-line AGN components, and other extended features.

Line identification combines automated template cross-correlation with photometric redshift constraints and extensive visual inspection to account for source confusion inherent in slitless spectroscopy. We further quantify morphological consistency between emission-line structures and direct imaging using a Siamese neural network. At the same time, a large language model (LLM) integrates spectral, photometric, and morphological information to assist in line assignment and redshift determination. Additional filtering steps remove spurious detections (e.g., cosmic rays, grism artifacts, and contamination from overlapping spectra). The resulting catalog provides a robust and statistically well-characterized sample of emission-line sources across a wide redshift range, forming the basis for our subsequent H$_2$ selection.

We defined a set of bright H$_2$ ro-vibrational transitions, including H$_2$ 1–0 S(1), S(3), Q(1), Q(3), O(3), and O(5). These lines are among the strongest transitions expected under local thermal equilibrium (LTE) at temperatures of $\sim$1000 K. We constructed an H$_2$-luminous sample by requiring that at least one of these lines is detected with a signal-to-noise ratio (S/N) greater than 3 in the 1D extracted spectrum. Figure~\ref{fig:ID204045} shows one example of an H$_2$ emitter with ID 204045. For each candidate, we then performed forced measurements of additional H$_2$ ro-vibrational lines on the 2D continuum-subtracted spectra, adopting the same spatial aperture defined by the primary detected bright line. This approach enables consistent flux extraction for weaker transitions and allows us to probe the full set of accessible H$_2$ lines. We retained sources with reliable detections (2D S/N $>$ 3) after visual inspection of the spectra and removal of contamination sources. 

\begin{figure}[t]

\centering
\includegraphics[width=0.49\textwidth]{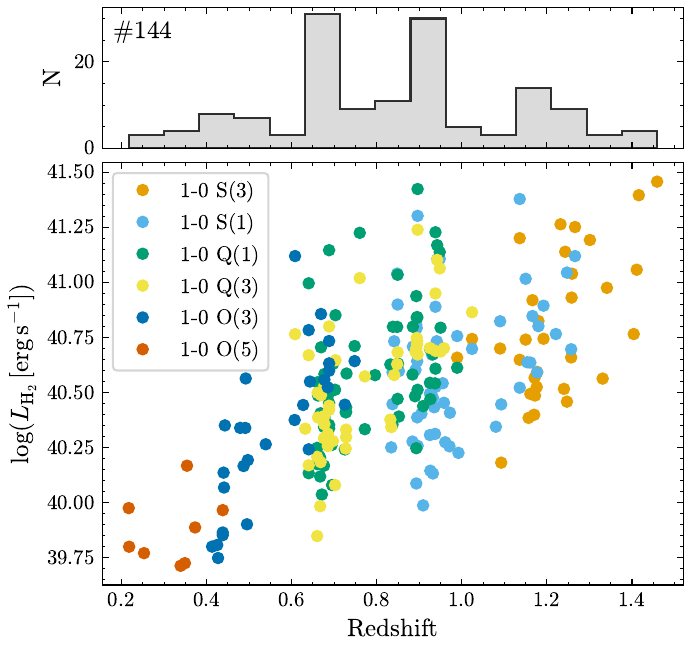}
\caption{H$_2$ luminosity vs. redshift distribution. The upper panel shows the redshift distribution of all 144 H$_2$ emitters. Note that some H$_2$ emitters are detected with multiple H$_2$ transitions.}
\label{fig:z-L}
\end{figure}

In total, our final sample consisted of 144 H$_2$-emitting galaxies spanning a redshift range of $z = 0.217$–1.459. Among them, 15 galaxies have one emission line in the F444W spectrum but have a reliable redshift based on the public optical spectra. The remaining 129 galaxies have at least two emission lines (H$_2$ or other lines). 56 galaxies have one H$_2$ line ($N_{\rm{lines}}=1$), and 88 galaxies have two or more H$_2$ lines ($N_{\rm{lines}}\geq2$). In summary, we have 58, 32, 67, 49, 29, and 8 samples with H$_2$ 1–0 S(1), S(3), Q(1), Q(3), O(3), and O(5). Figure~\ref{fig:z-L} shows the redshift and H$_2$ luminosity distribution of our sample. 

\subsection{Optical Spectra in COSMOS}

The extensive multi-wavelength coverage in the COSMOS field has enabled numerous spectroscopic follow-up programs. To characterize the nature of our sample, we utilize publicly available optical spectra from three major surveys: zCOSMOS \citep[][PI: S. Lilly]{2007Lilly,2009Lilly}, 10K-DEIMOS \citep{2018Hasinger}, and DESI-EDR \citep{2024DESI}. For DESI, the spectral resolutions of its three optical channels (blue: 3600-5900$\rm{\AA}$, green: 5660-7220$\rm{\AA}$, and red: 7470-9800$\rm{\AA}$) are R$\sim$2700, R$\sim$4200, and R$\sim$4600, respectively \citep{2024DESI}. The spectral resolutions of zCOSMOS and DEIMOS are $R \sim 600$ and $R \sim 5500$, respectively. In total, our sample includes 57 sources with DESI spectra, 82 with zCOSMOS, and 27 with DEIMOS coverage. These datasets provide the necessary emission-line measurements for optical diagnostic analyses.

\section{Results} 
\label{sec:H02}

\subsection{Multi-wavelength SEDs}

\begin{figure}
\centering
\begin{minipage}{0.45\textwidth}
  \centering
  \includegraphics[width=\textwidth]{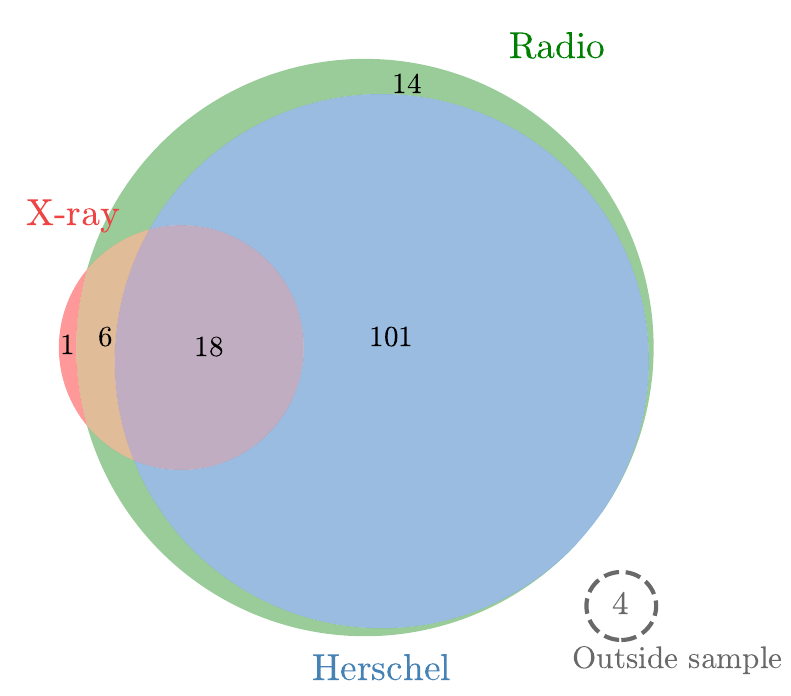}
\end{minipage}
\caption{
Venn diagram of our H$_2$ emitters showing detections in radio, X-ray, and Herschel single-band FIR emission. The gray dashed circle highlights the four H$_2$ emitters without detections in these bands. Most H$_2$ emitters are detected in both radio and FIR. }
\label{fig:venn}
\end{figure}

To characterize the multiwavelength properties of the H$_2$ sample, we compile photometry from the X-ray to radio by cross-matching the F444W source positions with existing COSMOS catalogs. The UV-to–NIR photometry is taken from the photometric catalog \citep{2026arXiv261013920F}, which combines data from CFHT, HSC, HST, and JWST with consistent photometric measurements. For MIR to submillimeter photometry, we adopt the super-deblended catalog \citep{superdeblended}. We match our sample to the super-deblended catalog using a radius of 1$''$, since the catalog provides flux measurements based on prior positions from higher-resolution $K_s$ imaging. The Venn diagram of our H$_2$ sample is shown in Figure~\ref{fig:venn}. 

For radio data, we cross-match the H$_2$ sample with several deep radio survey catalogs, including the VLA-COSMOS 1.4 GHz Large Project, the VLA 3 GHz survey, COSMOS-XS survey, and the MeerKAT MIGHTEE survey \citep{vla1.4ghz,vla3ghz,2021ApJvan}, as well as the MHz catalog from superMIGHTEE \citep{2025ApJ...991....9L}. In total, 139 sources (96.5\%) have at least one radio counterpart. The numbers of detections in the four individual catalogs are 88, 128, 136, and 113, respectively. This detection fraction is significantly higher than that of normal galaxy populations (typically $\sim$10\%), highlighting the strong association between H$_2$ emission and radio activity. For X-ray data, we match with the point source catalogs of the XMM-Newton Wide-Field Survey and the Chandra COSMOS Legacy Survey \citep{2010Brusa,2016Marchesi}, and obtain 24 and 14 objects, respectively. The total number of matched objects is 25 (17.4\%). The matching radius of the radio and X-ray catalog is 2$''$. 
 
\begin{figure}[t]
\centering
\begin{minipage}{0.49\textwidth}
  \centering
  \includegraphics[width=\textwidth]{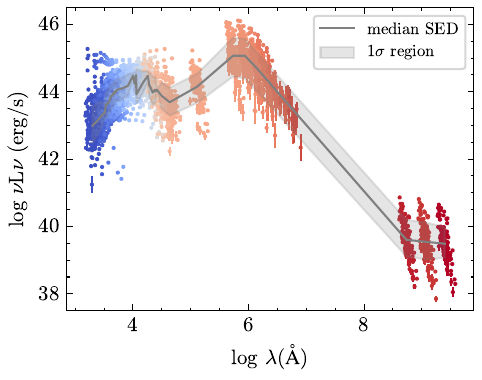}
\end{minipage}
\caption{UV–radio SED of the full H$_2$ sample. The gray solid curve shows the median SED, while the gray shaded region indicates the 16th–84th percentile range. The close similarity among individual SEDs suggests a common physical origin for the H$_2$ emitters.
}
\label{fig:SED}
\end{figure}

We estimate the total infrared luminosity ($L_{\rm{TIR}}$; $8$–$1000\,\mu$m) for 119 sources with at least one band from Herschel detection with S/N$>$3, and 114 sources fall in LIRGs and ULIRGs regimes, with $L_{\rm{TIR}} > 10^{11}\,L_\odot$. Five sources have marginally detected $L_{\rm{TIR}}\sim10^{10.5-10.9}\,L_\odot$. 80\% of H$_2$-luminous sources show bright FIR luminosity satisfying the (U)LIRGs definition. Using the compiled photometry, we construct SEDs for the full sample in Figure~\ref{fig:SED}. The SEDs exhibit broadly similar shapes, characterized by strong FIR emission. Hence, the H$_2$-bright sources are preferentially associated with systems exhibiting strong infrared and radio emission, suggesting a close connection between enhanced hot molecular gas emission and intense dusty star formation and energetic activity.

\subsection{The infrared–radio correlation}
  
We characterize the radio properties of the H$_2$ sample using multi-frequency observations available in the COSMOS field. For sources detected in at least two radio bands, we estimate the radio spectral index assuming a power-law spectrum, $S_\nu \propto \nu^{\alpha}$. We then derive rest-frame 1.4 GHz radio luminosities ($L_{1.4\,\mathrm{GHz}}$) by applying K-corrections based on the measured spectral index when available, or adopting a median spectral index of $\alpha \sim 0.75$.

One diagnostic of the relative radio contributions of star formation and AGN activity is the infrared–radio correlation \citep[IRRC; e.g.,][]{1985deJong}, which links the FIR and radio continuum emission in star-forming galaxies. The IRRC is commonly parameterized by
\begin{equation}
q_{\mathrm{IR}} = \log \left(\frac{L_{\mathrm{TIR}}[\mathrm{W}]}{3.75 \times 10^{12}\ \mathrm{Hz}}\right) - \log \left(L_{1.4\,\mathrm{GHz}}\,[\mathrm{W\, Hz^{-1}}]\right).
\end{equation}

\begin{figure}[t]
\centering
\includegraphics[width=0.49\textwidth]{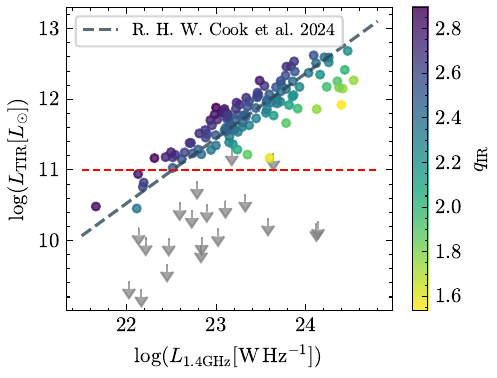}
\caption{Total infrared luminosity vs. radio luminosity for the full sample, color-coded by $q_{\rm IR}$. The gray triangles indicate sources without Herschel detections, shown as the upper limits. The blue dashed line represents the IRRC for star-forming galaxies \citep{2024Robin}. The red dashed line marks the luminosity threshold of LIRGs at $L_{\rm IR} = 10^{11}\,L_\odot$.}
\label{fig:TIR-Lradio}
\end{figure}

Star-forming galaxies typically follow a tight IRRC with $q_{\mathrm{IR}} \sim 2.6$, while radio-excess sources (i.e., $q_{\mathrm{IR}}<1.68$) are commonly associated with AGN-driven radio emission. 
Some recent work has argued that the IRRC evolves primarily with stellar mass, with more massive galaxies showing a systematically lower $q_{\rm{IR}}$ due to less energy loss of cosmic-ray electrons (CRe) \citep{2021A&A...647A.123D,2023A&A...679A..47S,2024ApJ...975...15Y}. As shown in Figure~\ref{fig:TIR-Lradio}, we present the IRRC relation for the H$_2$ sample. 114 sources with Herschel detections are shown as the filled circles color-coded with $q_{\rm{IR}}$, and the remaining 25 sources are shown as the gray triangles with upper limits. We find that H$_2$ sources with Herschel detections are broadly consistent with the IRRC of star-forming galaxies \citep{2024Robin}. In contrast, sources with low $q_{\rm IR}$ values and those with upper limits deviate from the IRRC, likely due to an enhanced radio-emission fraction associated with AGN activity or less CRe energy losses in massive galaxies ($M_* \gtrsim 10^{10.5}\,\rm{M_\odot}$).

\subsection{H$_2$ excitation temperature and hot H$_2$ mass}


To characterize the excitation of hot molecular gas, we construct H$_2$ excitation diagrams for sources with multiple detected H$_2$ emission lines in the 2D spectra. We select objects with more than two lines covering different upper-level energies ($N_{\rm{fit}}\geq2$), yielding a sample of 70 sources. Restricting to systems with more than three such lines ($N_{\rm{fit}}\geq3$) reduces the sample to 22 objects. We then calculate the H$_2$ excitation temperature for these 70 sources. For each transition, the column density of the upper level is derived from the measured line flux assuming optically thin emission, $N_{i}/g_{i} \propto F_{i}\lambda_{i}/(A_{i}g_{i})$, where $F_i$ and $\lambda_i$ are the line flux and wavelength, $A_i$ is the Einstein coefficient, and $g_i$ is the statistical weight of the upper level. Under the assumption of LTE, the level populations follow a Boltzmann distribution, $N_{i}/g_{i} \propto \exp[-E_{i}/(k_{B}T_{\rm{exc}})] $, where $E_i$ is the upper-level energy,  $T_{\rm exc}$ is the excitation temperature, and $k_{B}$ is the Boltzmann constant. Hence, the excitation temperature can be calculated by the ratio of the two H$_2$ line fluxes using
\begin{equation}
    T_{\mathrm{exc}}=-\frac{E_i-E_j}{k_{B} \ln \left(\left(\frac{F_i \lambda_i}{A_i g_i}\right) /\left(\frac{F_j \lambda_j}{A_j g_j}\right)\right)}
\end{equation}
For sources with more than three H$_2$ lines, we derive $T_{\rm exc}$ by fitting a linear regression in the excitation diagram.

\begin{figure}[t]
\centering
\begin{minipage}{0.49\textwidth}
  \centering
  \includegraphics[width=\textwidth]{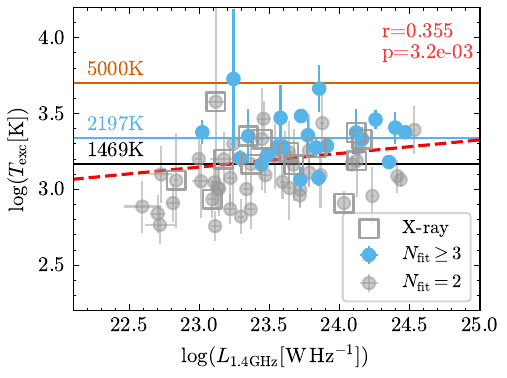}
\end{minipage}
\caption{
Excitation temperature vs. radio luminosity for 70 H$_2$ emitters. The blue circles denote 22 sources with at least three H$_2$ lines, while the gray circles represent the remaining 48 sources with two detected H$_2$ lines. The X-ray-detected sources are marked with the open squares. The red dashed line shows the best-fit linear relation for the full sample with $N_{\rm fit}\geq$2 H$_2$ lines.
}
\label{fig:temp}
\end{figure}

\begin{figure}[t]
\centering
\begin{minipage}{0.49\textwidth}
  \centering
  \includegraphics[width=\textwidth]{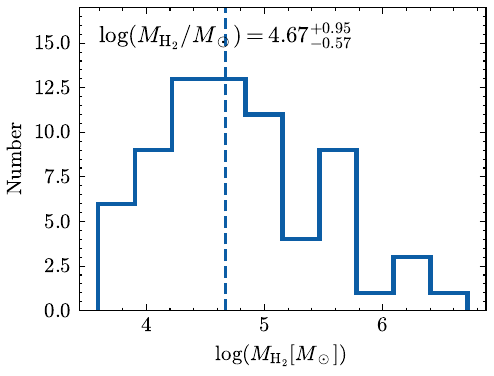}
\end{minipage}
\caption{
Distribution of the hot molecular gas masses estimated from the H$_2$ emission lines for the 70 H$_2$ emitters. The dashed vertical line marks the median value of $\log(M_{\rm H_2}/M_\odot)=4.67^{+0.95}_{-0.57}$, where the uncertainties indicate the 16th and 84th percentiles.
}
\label{fig:H2-mass-histogram}
\end{figure}

The resultant excitation temperature distribution is shown in Figure~\ref{fig:temp}. We find a median excitation temperature of $T_{\rm exc} \sim 1469\,\mathrm{K}$ for the the 70 sources with $N_{\rm fit} \geq 2$, and $T_{\rm exc} \sim 2197\,\mathrm{K}$ for the subsample of 22 sources with $N_{\rm fit} \geq 3$. The higher temperature in the latter sample reflects both the improved constraints provided by additional H$_2$ transitions and a selection effect toward sources with stronger high-excitation H$_2$ emission. Except for 3 sources showing clear non-linear excitation diagrams, most of the derived excitation temperatures are below the dissociation temperature of H$_2$ ($\sim4000$–$5000,\mathrm{K}$; \citealt{1999Hollenbach}), which also corresponds to the high excitation regime often observed in UV-pumped H$_2$ emission.

We estimate the temperature uncertainties through standard error propagation based on the slope of the linear regression. As expected, temperatures derived from $N_{\rm fit}=2$ exhibit larger uncertainties compared to those constrained by three or more lines. For the $N_{\rm fit} \geq 3$ subsample, the excitation is generally consistent with thermal processes, except for two sources that approach the dissociation threshold. We find a weak correlation between excitation temperature and radio luminosity ($r\sim 0.355$ and $p \sim 10^{-3}$). This weak correlation is driven primarily by the two-line subsample, while sources with more robust multi-line excitation constraints show systematically higher and radio-independent temperatures. This suggests that radio activity may increase the occurrence of highly excited molecular gas, but does not continuously raise the characteristic temperature of the hot H$_2$ component.

We also estimate the mass of the hot molecular gas traced by the H$_2$ emission lines using
\begin{equation}
M_{\mathrm{H}_2}=\frac{2 m_{\mathrm{p}} F_{i}4\pi D_{L}^{2}}{f_{i} A_{i} h \nu_{i}},
\end{equation}
where $m_{\mathrm{p}}$ is the proton mass, $f_{i}$ is the fractional population of the upper energy level, $h$ is the Planck constant, and $\nu_{i}$ is the photon frequency. We adopt the transition probabilities from \cite{Roueff2019}. Under the assumption of LTE, the fractional population of the upper level is determined by the excitation temperature through the Boltzmann distribution. For example, for $T_{\rm exc}\sim1500\,\mathrm{K}$, we obtain $f_{\nu=1, \mathrm{~J}=3}\sim0.007$ with H$_2$ 1-0 S(1). We estimate $M_{\mathrm{H}_2}$ from the available transitions and adopt the estimate from the highest-priority transition, following the order H$_2$ 1--0 S(1), Q(1), S(3), Q(3), O(3), and O(5). The mass estimates obtained from different transitions are generally consistent. The mass of the hot molecular gas derived from different transitions is consistent. The resulting distribution of hot molecular gas masses for the 70 H$_2$ emitters is shown in Figure~\ref{fig:H2-mass-histogram}, with a median value of $10^{4.67}\,M_\odot$.



\subsection{Strong H$_2$-radio association}

\begin{figure*}
\centering
\includegraphics[width=0.9\textwidth]{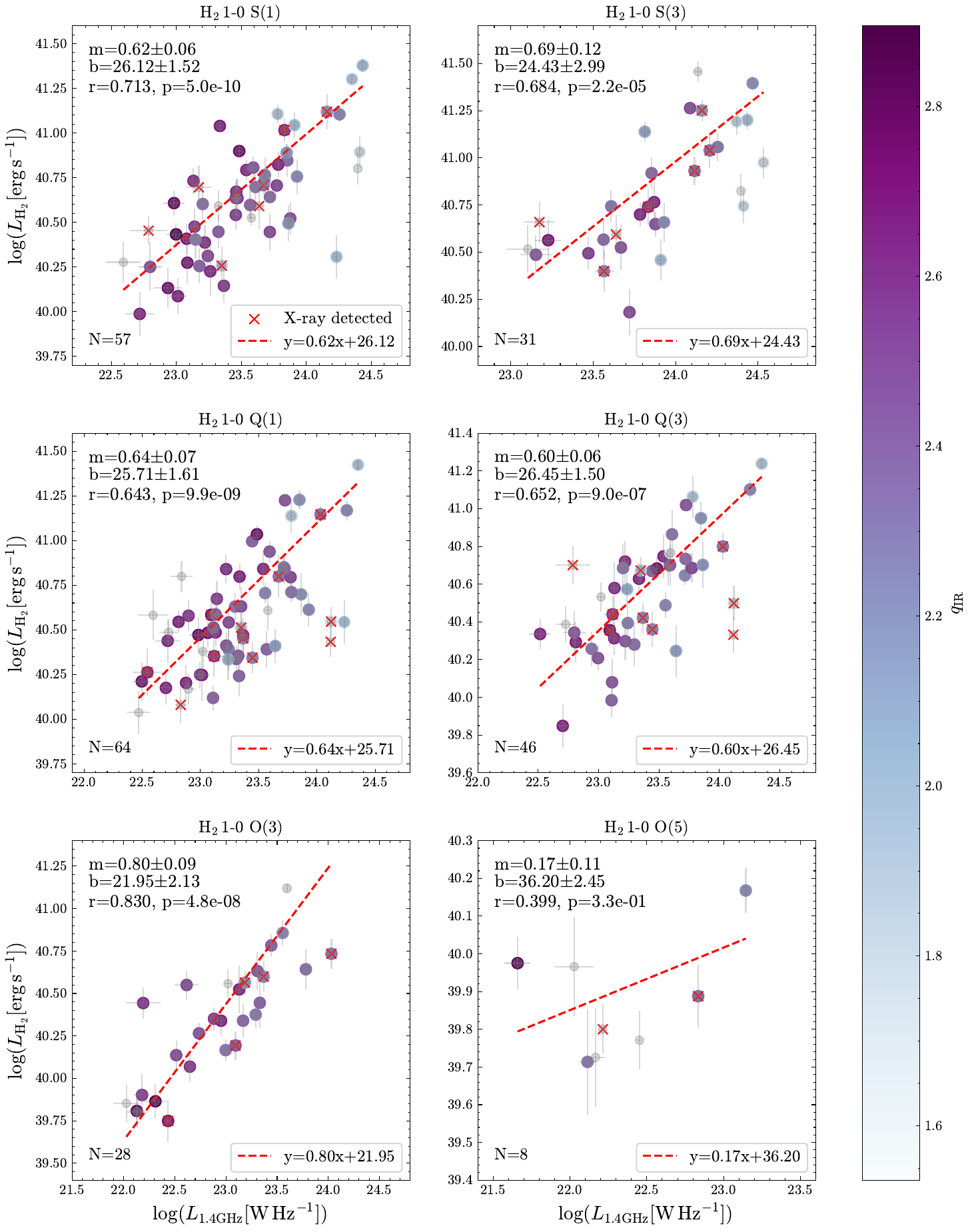}
\caption{H$_2$ 1--0 S(1), S(3), Q(1), Q(3), O(3), and O(5) line luminosities as a function of the rest-frame 1.4 GHz luminosity, color-coded by $q_{\rm IR}$. The X-ray–detected sources are marked with red crosses. The red dashed line shows the linear regression between the 1.4 GHz and H$_2$ luminosities. The gray data points represent radio H$_2$ emitters without Herschel detection. }
\label{fig:H$_2$-radio}
\end{figure*}

We find that 96.5\% of the H$_2$ sample has radio counterparts, enabling a robust association between hot molecular gas and radio emission. In Figure~\ref{fig:H$_2$-radio}, we compare the luminosities of the H$_2$ 1–0 S(1), S(3), Q(1), Q(3), O(3), and O(5) transitions with the rest-frame 1.4 GHz radio luminosity. The data points are color-coded by their $q_{\mathrm{IR}}$ values, and the X-ray–detected sources are marked with the red crosses. All H$_2$ transitions except H$_2$ 1–0 O(5) exhibit statistically significant positive correlations with radio luminosity. The H$_2$ 1–0 O(5) transition lies close to prominent PAH features, and the small sample size for this line ($N=8$) likely reduces the statistical significance of the correlation. The Pearson correlation coefficients in Table~\ref{tab:regression} indicate strong significance with $p \lesssim 10^{-5}$ for all transitions except H$_2$ 1–0 O(5). The consistency among different transitions indicates that the observed trend is not driven by a single line, but reflects a general connection between the hot H$_2$ emission and radio continuum.

We perform least-squares fits in the logarithmic space to quantify the relationship between the H$_2$ line luminosity and radio luminosity, adopting the form
\begin{equation}
\label{eq:h2-radio}
\log( L_\mathrm{H_2}/\mathrm{erg\,s^{-1}})= m \log( L_{1.4\mathrm{GHz}}/\mathrm{W\,Hz^{-1}})+b.
\end{equation}
The best-fit parameters (with $1\sigma$ uncertainties) are shown in Figure~\ref{fig:H$_2$-radio} and Table~\ref{tab:regression}. 
The uncertainties on the slope and intercept are estimated following \citet{Hogg2010}, under the assumption that the measurement uncertainties in $\log L_{\mathrm{H_2}}$ are Gaussian distributed. The tight correlations and high detection fraction together suggest a close physical connection between radio emission and the processes responsible for exciting hot molecular hydrogen. We will further discuss this relation in Section~\ref{sec:discussion}.

\begin{table}[t]
\centering

\caption{Fitting parameters between H$_2$ and radio luminosity and Pearson correlation coefficients for six H$_2$ transitions}
\begin{tabular}{lccccc}
\hline
\hline
Transitions & $m$ & $b$ & $r$ & $p$ \\
\hline
H$_2$ 1--0 S(1) & $0.62 \pm 0.06$ & $26.12 \pm 1.52$ & 0.713 & $4.99 \times 10^{-10}$ \\
H$_2$ 1--0 S(3) & $0.69 \pm 0.12$ & $24.43 \pm 2.99$ & 0.684 & $2.22 \times 10^{-5}$ \\
H$_2$ 1--0 Q(1) & $0.64 \pm 0.07$ & $25.71 \pm 1.61$ & 0.643 & $9.85 \times 10^{-9}$ \\
H$_2$ 1--0 Q(3) & $0.60 \pm 0.06$ & $26.45 \pm 1.50$ & 0.652 & $8.95 \times 10^{-7}$ \\
H$_2$ 1--0 O(3) & $0.80 \pm 0.09$ & $21.95 \pm 2.13$ & 0.830 & $4.75 \times 10^{-8}$ \\
H$_2$ 1--0 O(5) & $0.17 \pm 0.11$ & $36.20 \pm 2.45$ & 0.399 & $3.27 \times 10^{-1}$ \\
\hline
\end{tabular}
\label{tab:regression}
\end{table}
\subsection{SED modeling and extinction correction}

We perform SED modeling with \textsc{cigale} \citep{Boquien_2019} to study stellar populations for all H$_2$ galaxies following some local (U)LIRGs studies \citep[e.g.,][]{2024Dey}. We add an additional $5\%$ uncertainty in quadrature to all photometric flux uncertainties. We assume a commonly used, delayed star formation history (SFH; \texttt{sfhdelayed}), where the star formation rate $\mathrm{SFR} \propto te^{-t/\tau}$ and $\tau$ can vary in a wide range from 300\,Myr to 1\,Gyr. An optional late starburst is allowed in the latest 80–150\,Myr, which can contribute 0--60\% of the total stellar mass. The age of the main stellar population in the galaxy ranges from 1\,Gyr to 3.5\, Gyr. We use \citet{Bruzual_2003} stellar population synthesis models and adopt the \citet{Chabrier_2003} initial mass function (IMF). We also fix a metallicity of 0.02 $Z_\odot$ and an ionization parameter ($\log U$) of $-3.0$. We adopt the Milky Way extinction curve \citep{CCM89} for nebular emission and the modified \citet{Calzetti_2000} attenuation curve for the UV continuum \citep[e.g.,][]{Buat_2018}.
The color excess of the nebular lines is allowed between $E(B-V)_{\mathrm{line}}=0$--$2$ mag. We allow the ratio between $E(B-V)_{\mathrm{line}}$ and the color excess of the stellar continuum $E(B-V)_{\mathrm{stellar}}$ to vary between 0.44 and 0.7 \citep{Calzetti_2000}. For dust modeling, we use the \texttt{themis} model from \cite{2017Jones} with the mass fraction of hydrocarbon solids ranging from 0.02 to 0.1, minimum radiation field from 1 to 25, and the fraction illuminated from Umin to Umax from 0.02 to 0.1. For the AGN, we use AGN models from \cite{Stalevski2016}. The radio model includes galaxy synchrotron emission and AGN. The detailed model parameter grid is shown in Table~\ref{tab:CIGALE} in Appendix~\ref{sec:alllist}.


CIGALE estimates a comprehensive suite of physical parameters for each galaxy through a Bayesian-like analysis. In this work, we report the median values and associated uncertainties of the dust attenuation, $E(B-V)$, as well as other key stellar population properties, including SFR and stellar mass. The $E(B-V)$ values of the H$_2$ sample mainly range from 0.18 to 1.40. The median specific star formation rate (sSFR) is approximately $10^{-9.5}\,\mathrm{yr^{-1}}$, spanning a range from $10^{-10.0}$ to $10^{-8.3}\,\mathrm{yr^{-1}}$. This indicates that the H$_2$ sample predominantly consists of star-forming galaxies, with a small fraction reaching starburst-level activity.

\subsection{Star formation rates}

SFRs can be estimated using a variety of multi-wavelength tracers, spanning UV to radio wavelengths, including recombination lines (e.g., H$\alpha$, [O II]), UV continuum emission, FIR luminosity, radio continuum emission, and SED modeling. In dusty systems such as (U)LIRGs, UV and optical tracers are strongly affected by extinction, making FIR and radio luminosities the most commonly adopted SFR indicators for (U)LIRGs.

The widely used radio-based SFR calibration relies on the tight IRRC \citep{1985deJong,1985ApJ...298L...7H}, under the assumption that the radio emission is predominantly powered by star formation. However, this assumption may not hold in the presence of AGN. In particular, radio-quiet AGN can still follow the IRRC \citep[e.g.,][]{2010Moric}, while both their IR and radio emission are significantly contaminated or even dominated by AGN-related processes \citep[e.g.,][]{2016Zakamska}. Therefore, it is essential to compare SFR estimates derived from different tracers to assess their reliability, especially in AGN-hosting systems.

\begin{figure}[t]
\centering
\begin{minipage}{0.49\textwidth}
  \centering
  \includegraphics[width=\textwidth]{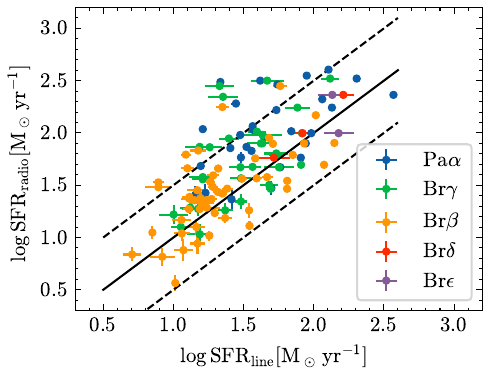}
\end{minipage}
\hfill
\begin{minipage}{0.48\textwidth}
  \centering
  \includegraphics[width=\textwidth]{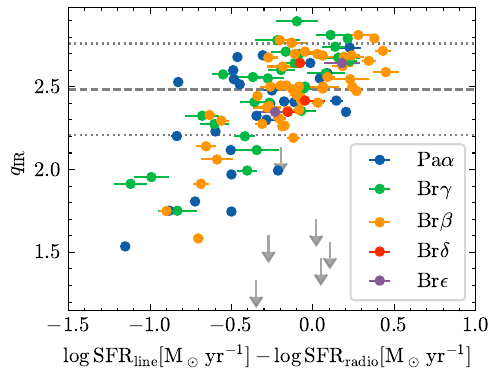}
\end{minipage}
\caption{
Upper panel: comparison between dust-corrected SFRs derived from hydrogen recombination lines in the F444W spectra and from 1.4 GHz radio luminosity using the calibration of \cite{2024Robin} with 100\,Myr timescale. The black solid line indicates the 1:1 relation, and the dashed lines mark deviations of $\pm$0.5 dex. Lower panel: difference between SFRs from recombination lines and radio emission as a function of $q_{\rm IR}$. The gray dashed and dotted lines indicate the median $q_{\rm IR}$ and its 1$\sigma$ uncertainty. The discrepancy between the two SFR estimators increases toward stronger radio-excess systems, exceeding 0.5 dex in $q_{\rm IR}<2.2$ sources.}
\label{fig:SFR_comparison}
\end{figure}

\begin{figure}[t]
\centering
\includegraphics[width=0.49\textwidth]{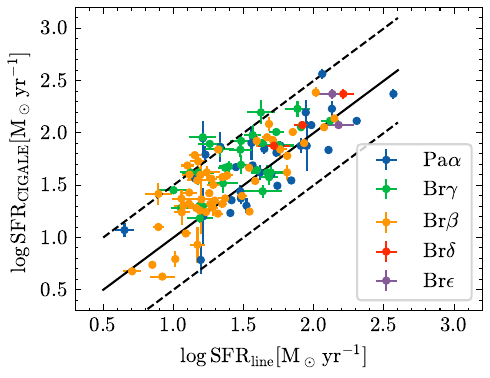}
\caption{Comparison between dust-corrected SFRs derived from hydrogen recombination lines in the F444W spectra and from SED modeling with CIGALE. The black solid line indicates the 1:1 relation, and the dashed lines mark deviations of $\pm$0.5 dex. The two SFR estimates are broadly consistent, agreeing within $\sim$0.5 dex.}
\label{fig:SFR_q}
\end{figure}

We compare SFRs derived from hydrogen recombination lines in the F444W spectra with those estimated from radio continuum emission and SED modeling. The recombination-line SFRs are computed by converting the Pa$\alpha$, Br$\beta$, Br$\gamma$, Br$\delta$, and Br$\epsilon$ luminosities to H$\alpha$ luminosity under the assumption of Case B recombination, and then applying the calibration of \citet{Kennicutt2012}. We apply the extinction corrections from the SED modeling to the recombination-line SFRs. The number of SFR measurements derived from recombination lines is 32, 52, 31, 3, and 2 for Pa$\alpha$, Br$\beta$, Br$\gamma$, Br$\delta$, and Br$\epsilon$, respectively. The radio SFRs are derived using 1.4\,GHz radio luminosity calibrations \citep{2024Robin}, assuming that the emission is dominated by star formation and the SFR is the average over 100\,Myr. In Figure~\ref{fig:SFR_comparison}, 21 sources have radio-derived SFRs more than 0.5 dex higher than recombination-line SFRs, and the remaining are consistent within 0.5 dex. These radio-SFR excess sources are mainly with $q_{\rm IR}<2.2$. We also compare the recombination-line SFRs with SED modeling-derived SFRs in Figure~\ref{fig:SFR_q}, and they are all consistent within 0.5 dex. In summary, we find that the SFRs of the H$_2$ sample mainly range from 4 to 350 $\rm{M}_\odot\,\mathrm{yr^{-1}}$. We compare the recombination-line SFRs with the star-forming main sequence from \citet{2012Whitaker}, accounting for the redshift and stellar mass from SED modeling. The majority of the sample lies within 0.5 dex of the main sequence, suggesting that most H$_2$-luminous galaxies are consistent with actively star-forming systems.

\section{DISCUSSION}
\label{sec:discussion}

In this section, we investigate the connection between AGN and star-forming activity and their relation to H$_2$ excitation in our sample.

\subsection{AGN activity}

\begin{figure}
\centering
\begin{minipage}{0.49\textwidth}
  \centering
  \includegraphics[width=\textwidth]{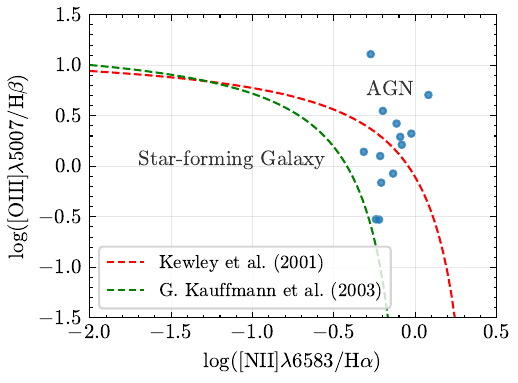}
\end{minipage}
\hfill
\begin{minipage}{0.48\textwidth}
  \centering
  \includegraphics[width=\textwidth]{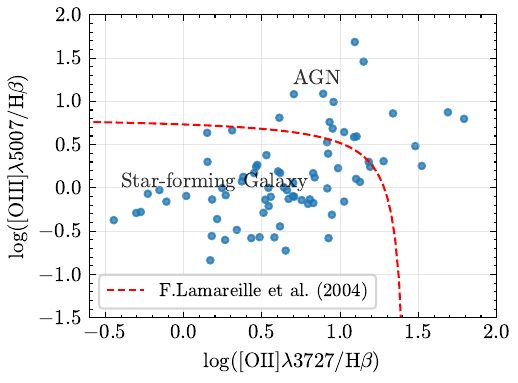}
\end{minipage}
\caption{Two BPT diagnostic diagrams for the H$_2$ sample with optical spectra from DESI, DEIMOS, and zCOSMOS. Upper panel: [NII]-based diagram using H$\alpha$/[NII]$\lambda6585$ versus [OIII]$\lambda5007$/H$\beta$. Lower panel: [OII]-based diagram using [OIII]$\lambda5007$/H$\beta$ versus [OII]$\lambda3727$/H$\beta$. These diagnostics reveal that a substantial fraction of H$_2$ emitters occupy the AGN regions.
}
\label{fig:BPT}
\end{figure}

We first identify AGN activity using X-ray luminosity and optical diagnostics with the BPT diagram. 25 H$_2$ emitters have X-ray detection with 2-10\,keV X-ray luminosity ranging from $10^{42.5-44.3}\, \rm{erg\,s^{-1}}$. A subsample of 13 sources at $z < 0.52$ has optical spectral coverage of the H$\alpha$/[NII]$\lambda6585$ and [OIII]$\lambda5007$/H$\beta$ emission lines, allowing us to construct the classical BPT diagram to separate star-forming galaxies from AGN. We measure emission-line fluxes by fitting Gaussian profiles to H$\alpha$, [NII]$\lambda6585$, H$\beta$, and [OIII]$\lambda5007$. For 78 sources with redshifts up to 1.0, we use diagnostic diagrams involving [OIII]$\lambda5007$/H$\beta$ and [O II]$\lambda3727$/H$\beta$ to constrain the ionization mechanism. However, this diagnostic is less reliable than the classical BPT diagram because [O II]$\lambda3727$ is more strongly affected by dust extinction and ionization parameter variations, introducing larger systematic uncertainties. We apply the dust extinction corrections from the SED modeling to the [O II]$\lambda3727$/H$\beta$ ratio. Figure~\ref{fig:BPT} shows two BPT diagnostic diagrams, with classification schemes adopted from \cite{2001Kewley,2003Kauffmann,2004Lamareille}.

\begin{figure}

\centering
\includegraphics[width=0.48\textwidth]{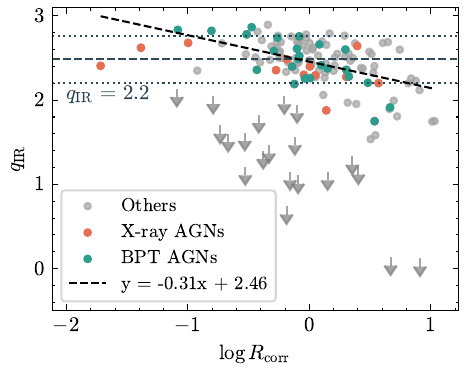}
\caption{$q_{\mathrm{IR}}$ vs radio loudness $R = f_{\mathrm{5\,GHz}}/f_{\mathrm{4400}}$ for the H$_2$ sample. The orange and green points denote the X-ray AGNs and BPT-selected AGNs, respectively. The H$_2$ emitters without Herschel detections are shown as the gray upper limits. The gray dashed and dotted lines indicate the median $q_{\rm IR}$ and its 1$\sigma$ dispersion. The anti-correlation between $q_{\rm IR}$ and $R$ is consistent with enhanced AGN-driven radio emission.}
\label{fig:qir-radio}
\end{figure}

We also derive radio loudness $R = f_{\rm{5GHz}}/f_{\rm{4400}}$ for the H$_2$ samples, where $f_{\rm{5GHz}}$ and $f_{\rm{4400}}$ are the rest-frame flux densities at 5$\,$GHz and $4400\,\rm{\AA}$ based on the extrapolation or interpolation of the radio and optical SEDs. Dust extinction corrections have been applied to $f_{\rm{4400}}$. We find an anti-correlation between $q_{\rm{IR}}$ and $R$ with a Pearson correlation coefficient of $r=-0.519\,(p=1.76\times 10^{-9})$, as shown in Figure~\ref{fig:qir-radio}. After dust correction, most H$_2$ samples are radio-quiet. Hence, we adopt $q_{\rm {IR}} <2.2$ (corresponding to a $1\sigma$ radio excess) with Herschel detection or $|\log (\rm{SFR}_{radio}/\rm{SFR}_{CIGALE})|>0.5$ as the criterion to identify radio-excess AGNs.

\begin{figure}

\centering
\includegraphics[width=0.48\textwidth]{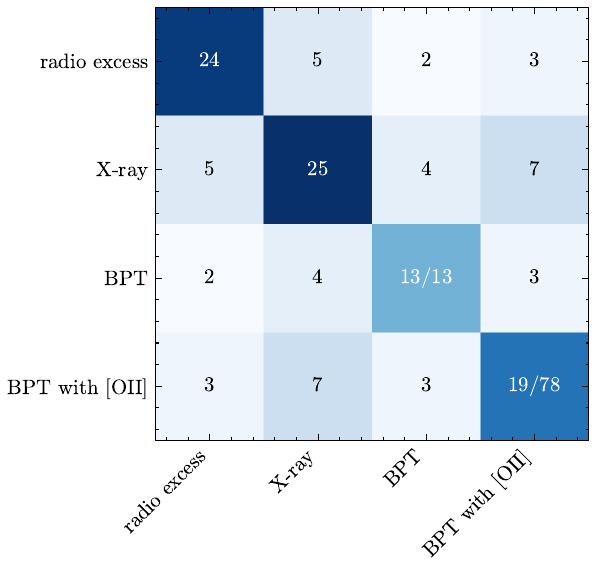}
\caption{AGN classification of the H$_2$ sample based on radio excess, X-ray detection, and two BPT diagnostics. In total, 62 H$_2$ emitters are identified as AGN by at least one method. The different diagnostics trace partially distinct AGN populations and together significantly increase the overall AGN identification completeness.}
\label{fig:agn}
\end{figure}

Combining X-ray luminosity, radio excess, and optical emission-line diagnostics, we identify at least 62 sources with clear evidence of AGN activity, as shown in Figure~\ref{fig:agn}. We do not use radio loudness as an AGN indicator, since dust obscuration can significantly affect the optical continuum and bias the inferred radio-to-optical ratios.

\subsection{NIR Diagnostics}

NIR emission-line ratios provide a diagnostic of gas excitation mechanisms in dusty galaxies. Using VLT/SINFONI integral-field spectroscopy, \citet{Colina2015} introduced a BPT-like diagram based on Br$\gamma$, [Fe II] 1.64\,$\mu$m, and H$_2$ 1--0 S(1) emission to distinguish between star formation, AGN activity, and supernova (SN)-dominated excitation in local (U)LIRGs. In this framework, young star-forming regions (ages $\lesssim 6$ Myr) are characterized by both low [Fe II]/Br$\gamma$ and H$_2$/Br$\gamma$ ratios. In contrast, AGN-dominated regions exhibit elevated [Fe II]/Br$\gamma$ ratios, and SN-dominated regions typically show intermediate H$_2$/Br$\gamma$ values between those of star-forming regions and AGN. Previous results indicate that the ionization and heating in local LIRGs arise from a combination of processes.

Empirically, star-forming regions with FUV-pumped fluorescence generally exhibit H$_2$/Br$\gamma \lesssim 0.4$–0.8 \citep[e.g.,][]{1994ApJ...421..115D,1988MNRAS.234P..29P,1998MNRAS.295...43D}, while higher values suggest increasing contributions from AGN activity or shocks \citep[e.g.,][]{1998Larkin,2005Rodr,2008Riffel,2013Riffel}. More recently, a study based on 453 hard X-ray–selected AGNs from the BASS survey proposed a refined classification, with H$_2$/Br$\gamma \lesssim 0.25$ for star-forming systems, H$_2$/Br$\gamma > 2$ for AGN-dominated sources, and intermediate values corresponding to composite (AGN+SF) regions \citep{Gillette2026}. The exact boundaries of this classification are not well defined and should be interpreted with caution.

\begin{figure}
\centering
\includegraphics[width=0.49\textwidth]{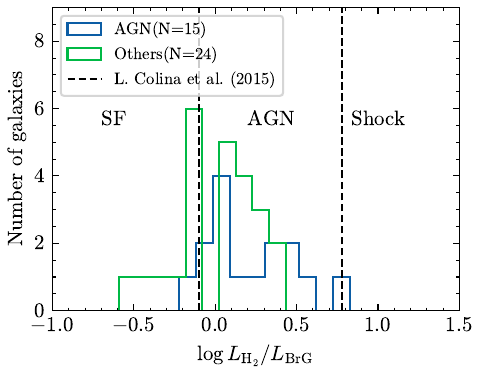}
\caption{
NIR diagnostic diagram showing the distribution of the H$_2$/Br$\gamma$ ratios for 39 sources, including 15 AGN and 24 other objects classified in this work. The black dashed lines indicate the boundaries between the star-forming, AGN, and shock-dominated excitation regimes, as defined by \citet{Colina2015}. 
}
\label{fig:H2/BrG}
\end{figure}

For our sample of 39 sources, we measure their H$_2$/Br$\gamma$ ratios using the H$_2$ 1–0 S(1) transition. For sources without direct Br$\gamma$ detection, we estimate the Br$\gamma$ luminosity by converting other hydrogen recombination lines under the assumption of Case B recombination. Figure~\ref{fig:H2/BrG} shows the resulting H$_2$/Br$\gamma$ distribution. We further compare the H$_2$/Br$\gamma$ ratios of H$_2$ emitters with and without AGN activity. The AGN subsample generally occupies the AGN regime defined by \citet{Colina2015}, with H$_2$/Br$\gamma > 0.4$. The substantial fraction of the remaining sources also exhibits elevated H$_2$/Br$\gamma$ ratios, but does not show significant AGN activity. This suggests that our sample does not fully satisfy previous classifications.

\subsection{H$_2$ Excitation and Connection to IR and Radio Emission}

The IRRC is believed to arise from the fact that both IR and radio continuum are from star formation. TIR emission predominantly traces recent star formation, arising from dust heated by UV radiation from massive ($>5\,\rm{M_\odot}$) O and early-type B stars and re-emitted in the infrared. The radio continuum emission at $\sim$1.4 GHz is primarily produced by synchrotron radiation from CRe, which are accelerated in supernova remnants associated with the deaths of more massive stars ($>8\,\rm{M_\odot}$). The NIR H$_2$ emission can be excited through multiple mechanisms, including UV heating in PDRs, X-ray/UV irradiation from AGN, and shocks driven by supernova remnants, AGN outflows/jets, or dynamical processes such as galaxy merging.

\begin{figure*}

\centering
\includegraphics[width=0.9\textwidth]{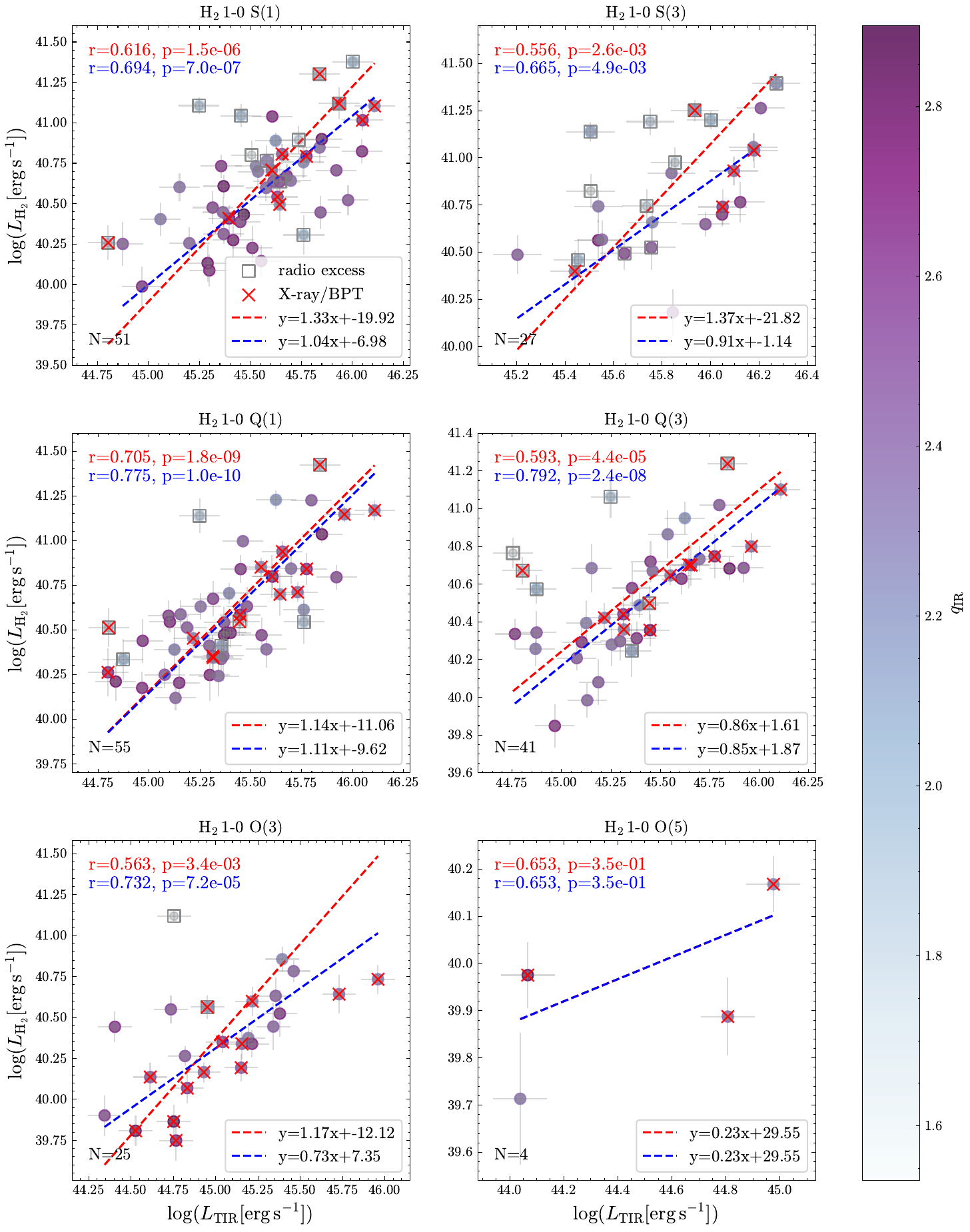}
\caption{H$_2$ 1--0 S(1), S(3), Q(1), Q(3), O(3), and O(5) line luminosities as a function of total infrared luminosity for our H$_2$ emitters with Herschel detections, color-coded by $q_{\rm IR}$. The radio-excess AGNs are marked with the gray open squares, while the X-ray–detected and BPT AGNs are marked with the red crosses. The red dashed lines show the linear regressions between $L_{\rm TIR}$ and the H$_2$ line luminosities for the full sample, while the blue dashed lines show the corresponding regressions excluding the radio-excess sources.}
\label{fig:H$_2$-TIR}
\end{figure*}

\begin{figure}
\centering
\includegraphics[width=0.49\textwidth]{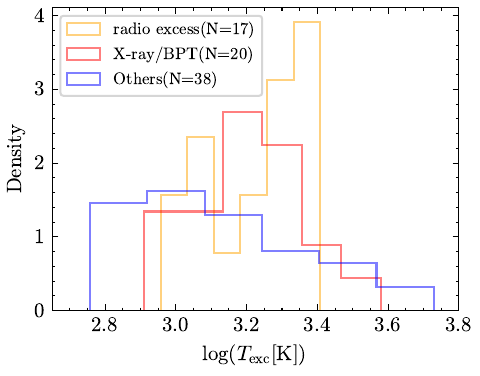}
\caption{
Histogram of excitation temperatures for 17 radio-excess AGNs (orange), 20 X-ray/BPT-detected AGNs (red), and 38 remaining sources (blue).}
\label{fig:Tex_hist}
\end{figure}

\begin{figure*}
\centering

\begin{minipage}{0.78\textwidth}
  \centering
  \includegraphics[width=\textwidth]{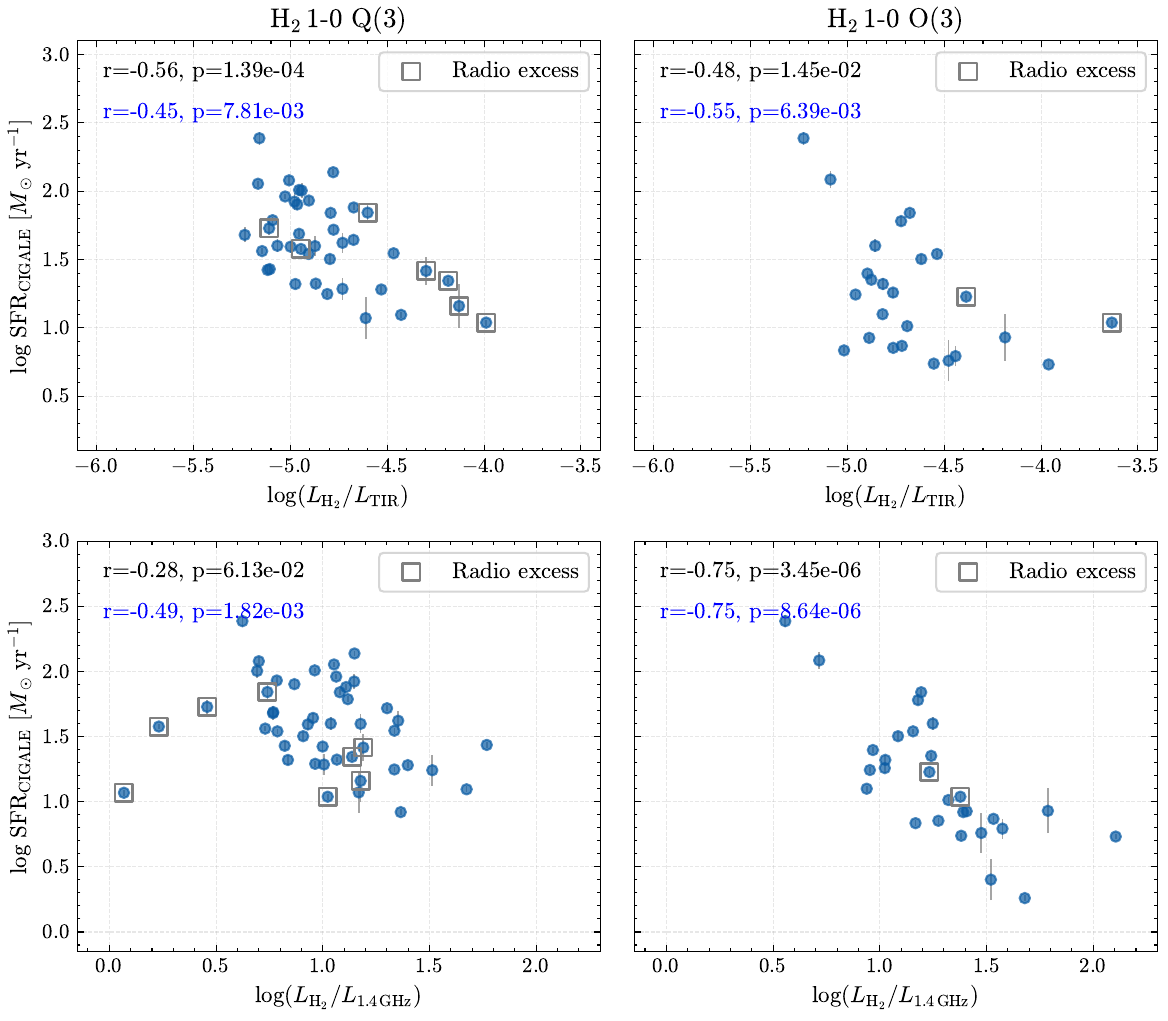}
\end{minipage}

\caption{SFR vs. $L_{\rm H_2}/L_{\rm TIR}$ and $L_{\rm H_2}/L_{1.4\,\rm GHz}$ for H$_2$ 1--0 Q(3) and H$_2$ 1--0 O(3), respectively. The Pearson correlation coefficients and corresponding $p$-values for the full sample and the sample excluding radio-excess sources are shown in black and blue, respectively. Gray open squares denote radio-excess sources.
}
\label{fig:SFR-H2}
\end{figure*}

In our H$_2$-luminous sample, we find strong positive correlations between the H$_2$ luminosities of the H$_2$ 1--0 S(1), S(3), Q(1), Q(3), and O(3) transitions and both radio and IR luminosities. Figure~\ref{fig:H$_2$-TIR} shows the relation between $L_{\rm H_2}$ and $L_{\rm TIR}$, with AGN classifications indicated (the X-ray/BPT AGNs as the red crosses and the radio-excess sources as the gray open squares). Note that here we only discuss the H$_2$ emitters with Herschel detection, because the large deblending uncertainty and shallow detection on Herschel images make a large uncertainty on $L_{\rm TIR}$ for non-detection objects. We perform linear regressions between $L_{\rm H_2}$ and $L_{\rm TIR}$ both for the full sample and for the sample excluding the radio-excess sources. We find no significant offset in $L_{\rm H_2}$ for the X-ray/BPT AGNs and star-forming galaxies, while nearly half of the radio-excess systems deviate toward enhanced $L_{\rm H_2}$ relative to the linear $L_{\rm H_2}$–$L_{\rm TIR}$ relation. The approximately linear $L_{\rm H_2}$--$L_{\rm TIR}$ relation after excluding the radio-excess sources suggests that this hot H$_2$ emission broadly scales with star formation activity. This trend suggests that H$_2$ emission in X-ray/BPT AGNs and star-forming galaxies is primarily powered by star formation-related processes, whereas radio-excess systems require an additional excitation channel, likely associated with jet- or wind-driven shocks.

The scaling relations between $L_{\rm H_2}$ and radio/infrared luminosities provide further insight into the excitation efficiency of hot H$_2$. Although $L_{\rm H_2}$ shows positive correlations with both $L_{1.4\,\rm GHz}$ and $L_{\rm TIR}$, the sublinear dependence of $L_{\rm H_2}$ on $L_{1.4\,\rm GHz}$ suggests that $L_{\rm H_2}/L_{1.4\,\rm GHz}$ decreases toward higher radio luminosities. In contrast, the near-linear $L_{\rm H_2}$--$L_{\rm TIR}$ relation indicates that $L_{\rm H_2}/L_{\rm TIR}$ remains approximately constant over the explored infrared luminosity range. These results suggest that star formation provides a baseline contribution to hot H$_2$ excitation, whereas the sublinear $L_{\rm H_2}$--$L_{1.4\,\rm GHz}$ relation may indicate that the coupling efficiency between radio-related energy output and hot molecular gas excitation varies with radio luminosity. At higher radio luminosities, an increasing fraction of the radio emission may not be effectively converted into hot H$_2$ emission, or the radio activity and molecular gas response may occur on different timescales.



We further examine the distribution of $T_{\rm exc}$ among radio-excess sources, X-ray/BPT AGNs, and the remaining galaxies. The median $T_{\rm exc}$ values are $1885^{+492}_{-690}\,\rm{K}$, $1561^{+683}_{-683}\,\rm{K}$, and $1224^{+1185}_{-455}\,\rm{K}$ for radio-excess sources, X-ray/BPT AGNs, and the remaining galaxies, respectively. Although no clear correlation is found between $T_{\rm exc}$ and the H$_2$ excitation efficiencies, parameterized by $L_{\rm H_2}/L_{1.4\,\rm GHz}$ and $L_{\rm H_2}/L_{\rm TIR}$, radio-excess and X-ray/BPT AGNs tentatively show higher excitation temperatures than the remaining galaxies (KS test, $p=0.13$ and $p=0.30$; Figure~\ref{fig:Tex_hist}). Among the H$_2$ emitters with $T_{\rm exc}$ estimation, 35/70 with $N_{\rm fit}\geq2$ and 10/22 with $N_{\rm fit}\geq3$ are AGNs. This may indicate that AGN-related processes could contribute to producing highly excited hot H$_2$, and the excitation temperature is regulated by physical conditions beyond AGN activity alone.

We also examine the dependence of $\rm{SFR}$ derived from the SED modeling on the H$_2$ excitation efficiencies. We find weak anti-correlations between the $\rm{SFR}$ and excitation efficiencies in all H$_2$ transitions with $r<0$, and only two transitions show statistically significant correlations in Figure~\ref{fig:SFR-H2}. This tentative trend suggests that ISM environments with enhanced hot H$_2$ emission may be associated with weak star formation activity. However, whether hot molecular gas excitation directly influences star formation or instead traces other processes, such as AGN feedback or host galaxy evolution, remains unclear. Radio-bright galaxies are often associated with higher stellar masses, more massive host halos, earlier-type, bulge-dominated systems, and lower SFRs. These correlations may introduce additional dependencies among SFR, radio activity, and molecular gas excitation. A detailed investigation of the coupled effects of AGN activity, host galaxy properties, and star formation is beyond the scope of this work and requires future studies with larger samples and additional diagnostics, e.g., outflows.

Overall, our results are consistent with a multi-channel excitation scenario in which both radiative and mechanical processes contribute to the observed H$_2$ emission. Currently, the absence of high signal-to-noise measurements of higher-excitation ro-vibrational lines (e.g., H$_2$ 2--1 S(1)) prevents a more definitive separation of UV radiation and shock excitation. In addition, the lack of spatially resolved spectroscopy limits our ability to disentangle AGN-related and star formation-related contributions to the hot molecular gas excitation. Future observations with higher spectral sensitivity and spatial resolution will be essential to constrain the relative contributions of different excitation mechanisms.

\section{SUMMARY}
\label{sec:Summary}

We have built a sample of 144 H$_2$-luminous galaxies that span a redshift range of 0.2 to 1.5. We constructed their multi-wavelength SEDs, studied their physical properties, and explored the mechanism for the H$_2$ emission. The main results are summarized below.
\begin{enumerate}

\item The H$_2$-luminous sample is predominantly composed of infrared-bright star-forming galaxies. Approximately 80\% of the sources satisfy the (U)LIRG criteria based on their FIR luminosities, and their SED shapes and SFRs are broadly consistent with star-forming galaxies (see Figures~\ref{fig:venn}, \ref{fig:SED} and \ref{fig:SFR_comparison}; Section~\ref{sec:H02}).

\item We identify 62 H$_2$-luminous sources with significant AGN activity based on radio excess, X-ray emission, and optical BPT diagnostics. Consistent results are obtained from NIR diagnostics using the H$_2$/Br$\gamma$ ratio for 39 sources (see Figures~\ref{fig:BPT} and \ref{fig:agn}; Section~\ref{sec:discussion}).

\item The H$_2$ luminosities of the H$_2$ 1--0 S(1), S(3), Q(1), Q(3), O(3), and O(5) transitions show strong positive correlations with both radio and infrared luminosities. The approximately linear $L_{\rm H_2}$--$L_{\rm TIR}$ relation after excluding radio-excess sources suggests that hot H$_2$ emission broadly scales with star formation activity, while radio-excess systems show enhanced $L_{\rm H_2}$ and likely require additional excitation mechanisms (see Figures~\ref{fig:H$_2$-radio} and \ref{fig:H$_2$-TIR}; Sections~\ref{sec:H02} and \ref{sec:discussion}).

\item The different scaling behaviors of hot H$_2$ with radio and infrared luminosities provide insights into the excitation efficiency. The sublinear $L_{\rm H_2}$--$L_{1.4\,\rm GHz}$ relation suggests that the coupling between radio-related energy output and hot molecular gas excitation varies with radio luminosity. This may reflect changes in energy coupling efficiency or timescale differences between radio activity and the molecular gas response (see Figures~\ref{fig:H$_2$-radio} and \ref{fig:H$_2$-TIR}; Sections~\ref{sec:H02} and \ref{sec:discussion}).

\item Radio-excess and X-ray/BPT AGNs tend to exhibit higher excitation temperatures than the remaining galaxies, although the difference is not statistically significant. No significant correlation is found between $T_{\rm exc}$ and the H$_2$ excitation efficiencies ($L_{\rm H_2}/L_{\rm TIR}$ and $L_{\rm H_2}/L_{1.4\,\rm GHz}$), suggesting that the excitation temperature and hot H$_2$ emission efficiency are regulated by different physical conditions (see Figures~\ref{fig:temp} and \ref{fig:Tex_hist}; Sections~\ref{sec:H02} and \ref{sec:discussion}).

\item We find tentative anti-correlations between SFR and H$_2$ excitation efficiencies, while statistically significant results are only in two transitions. This trend suggests that hot H$_2$-rich ISM environments may be associated with weak star formation activity (see Figure~\ref{fig:SFR-H2}; Section~\ref{sec:discussion}).

\end{enumerate}

\begin{acknowledgments}
We thank Shengzhe Wang and Siqi Liu for their discussion and suggestions. We acknowledge support from the National Science Foundation of China (12225301).
\end{acknowledgments}

\software{AstroML \citep{VanderPlas2012}, Jupyter notebooks \citep{Perez2007}, Matplotlib \citep{Hunter2007}, Numpy \& Scipy \citep{scipy}, SCIKIT-LEARN \citep{Pedregosa2011}.}

\vspace{5mm}


\appendix

\section{Measured Properties of the H$_2$ sample}\label{sec:alllist}
We present the list of input parameters used for CIGALE modeling in Table~\ref{tab:CIGALE}. We present the measured properties of our H$_2$ samples in Table~\ref{tab:H2}. The full table is available in machine-readable format online.


\begin{deluxetable*}{ccc}
\tablecaption{List of Input Parameters Used for CIGALE Modeling
\label{tab:CIGALE}}
\tablewidth{0pt}
\tablehead{
\colhead{Parameters} & \colhead{Symbol} & \colhead{Values}
}
\startdata
\multicolumn{3}{l}{\textbf{Delayed star formation history + optional exponential burst}} \\
\hline
e-folding time of the main stellar population [Myr] 
& $\tau_{\rm main}$ & 300, 500, 1000 \\
e-folding time of the late starburst population [Myr] 
& $\tau_{\rm burst}$ & 50 \\
Mass fraction of the late burst population 
& $f_{\rm burst}$ & 0.05, 0.2, 0.6 \\
Age of the main stellar population [Myr] 
& age & 1000,2000,3500 \\
Age of the late burst [Myr] 
& $\rm{age}_{\rm burst}$ & 80.0, 110, 150 \\
\hline
\multicolumn{3}{l}{\textbf{Stellar synthesis population \citep{Bruzual_2003}}} \\
\hline
Initial mass function 
& IMF & Salpeter (1955) \\
Metallicity 
& $Z$ & 0.02 \\
Separation age 
& -- & 1 Myr \\
\hline
\multicolumn{3}{l}{\textbf{Dust attenuation laws \citep{Calzetti_2000}}} \\
\hline
Color excess of young stars 
& $E(B-V)$ & 0.0, 0.4, 0.8, 1.2, 1.6, 2.0 \\
Reduction factor 
& $f_{\rm att}$ & 0.44, 0.7 \\
\hline
\multicolumn{3}{l}{\textbf{Dust grain model: THEMIS \citep{2017Jones}}} \\
\hline
Fraction of small hydrocarbon solids 
& $q_{\rm hac}$ & 0.02, 0.06, 0.10 \\
Minimum radiation field 
& $U_{\rm min}$ & 1, 10, 25 \\
Power-law index of radiation field 
& $\alpha$ & 2 \\
Fraction illuminated from $U_{\rm min}$ to $U_{\rm max}$ 
& $\gamma$ & 0.02, 0.06, 0.1 \\
\hline
\multicolumn{3}{l}{\textbf{Active nucleus nodel: SKIRTOR \citep{Stalevski2016}}} \\
\hline
Optical depth at 9.7 $\mu$m 
& $\tau_{9.7}$ & 3.0, 7.0 \\
Torus density radial parameter 
& $p_l$ & 1.0 \\
Torus density angular parameter 
& $q$ & 1.0 \\
Opening angle of torus [deg] 
& $oa$ & 40.0 \\
Outer-to-inner radius ratio 
& $R$ & 20.0 \\
Fraction of dust mass in clumps 
& $M_{\rm cl}$ & 0.97 \\
Inclination angle [deg] 
& $i$ & 30 (type 1), 70 (type 2) \\
AGN fraction 
& $f_{\rm AGN}$ & 0.05, 0.2, 0.4 \\
Polar dust extinction law 
& -- & SMC \\
$E(B-V)$ of polar dust 
& -- & 0.01, 0.5 \\
Temperature of polar dust [K] 
& -- & 100 \\
Emissivity index of polar dust 
& $\beta$ & 1.6 \\
\hline
\multicolumn{3}{l}{\textbf{Radio}} \\
\hline
FIR/radio parameter 
& $q_{\rm IR}$ & 2.6 \\
Star formation spectral slope 
& $\alpha_{\rm SF}$ & 1.7, 1.0, 0.5, 0.2 \\
Radio-loudness parameter 
& $R_{\rm AGN}$ & 0.1, 0.5, 10, 60, 200  \\
AGN spectral slope 
& $\alpha_{\rm AGN}$ &  1.7, 1.0, 0.5, 0.2 \\
\enddata

\end{deluxetable*}

\begin{deluxetable*}{cccc}
\tablecaption{Measured Properties of the H$_2$ galaxy \label{tab:H2}}
\tablewidth{20pt}
\tablehead{\colhead{Column} &\colhead{Format} &\colhead{Units} &\colhead{Description}}
\startdata
ID & int &  & Target ID in COSMOS2025 catalog \\
z & float &  & Spectroscopic redshift \\
F444W\_MAG & float & mag & Magnitude of JWST F444W band \\
F444W\_MAGERR & float & mag & Uncertainty of F444W\_MAG \\
H2\_LINES & array &  & Detected H$_2$ lines \\
LOGL\_H2 & array & $\mathrm{erg\ s^{-1}}$ & Logarithm of H$_2$ line luminosity \\
LOGL\_H2ERR & array & $\mathrm{erg\ s^{-1}}$ & Uncertainty of LOGL\_H2 \\
TEMP\_ex & float & K & Excitation temperature \\
TEMP\_exERR & float & K & Uncertainty of TEMP\_ex \\
LOGL\_1p4GHz & float & $\mathrm{erg\ s^{-1}}$ & Radio luminosity at 1.4 GHz \\
LOGL\_1p4GHzERR & float & $\mathrm{erg\ s^{-1}}$ & Uncertainty of LOGL\_1p4GHz \\
LOGL\_TIR & float & $\mathrm{erg\ s^{-1}}$ & Total infrared luminosity \\
LOGL\_TIRERR & float & $\mathrm{erg\ s^{-1}}$ & Uncertainty of LOGL\_TIR \\
LOGL\_Xray & float & $\mathrm{erg\ s^{-1}}$ & 2-10\,keV X-ray luminosity \\
LOGL\_XrayERR & float & $\mathrm{erg\ s^{-1}}$ & Uncertainty of LOGL\_Xray \\
qIR & float &  & Infrared-to-radio ratio parameter $q_{\mathrm{IR}}$ \\
AGN\_flag & int &  & Classification flag for AGN activity (1 or 0) \\
LINE\_NAME & str &  & Emission line name for SFR\\
SFR\_LINE & float & $\rm{M}_{\odot}\ \mathrm{yr^{-1}}$ & Star formation rate derived from line emission \\
SFR\_LINEERR & float & $\rm{M}_{\odot}\ \mathrm{yr^{-1}}$ & Uncertainty of SFR\_LINE \\
SFR\_CIGALE & float & $\rm{M}_{\odot}\ \mathrm{yr^{-1}}$ & Star formation rate derived from CIGALE SED fitting \\
SFR\_CIGALEERR & float & $\rm{M}_{\odot}\ \mathrm{yr^{-1}}$ & Uncertainty of SFR\_CIGALE \\
STELLAR\_MASS\_CIGALE & float & $\rm{M}_{\odot}\ \mathrm{yr^{-1}}$ & Stellar mass derived from CIGALE SED fitting \\
STELLAR\_MASSERR & float & $\rm{M}_{\odot}\ \mathrm{yr^{-1}}$ & Uncertainty of STELLAR\_MASS\_CIGALE \\
SFR\_RADIO & float & $\rm{M}_{\odot}\ \mathrm{yr^{-1}}$ & Star formation rate derived from radio luminosity \\
SFR\_RADIOERR & float & $\rm{M}_{\odot}\ \mathrm{yr^{-1}}$ & Uncertainty of SFR\_RADIO \\
EBV & float & mag & Dust reddening $E(B-V)$ from CIGALE SED fitting  \\
EBV\_ERR & float & mag & Uncertainty of EBV \\
\enddata
\tablecomments{This table is available in its entirety in machine-readable format.}
\label{tab:H2}
\end{deluxetable*}

\bibliography{sample631}{}
\bibliographystyle{aasjournalv7}

\end{document}